Dyadic:

A Scalable Platform for Human-Human and Human-AI Conversation Research

David M. Markowitz[1]

[1] Department of Communication, Michigan State University, East Lansing, MI 48824

**Abstract**

Conversation is ubiquitous in social life, but the empirical study of this interactive process has been thwarted by tools that are insufficiently modular and unadaptive to researcher needs. To relieve many constraints in conversation research, the current tutorial presents an overview and introduction to a new tool, Dyadic (https://www.chatdyadic.com/), a web-based platform for studying human-human and human-AI conversations using text-based or voice-based chats. Dyadic is distinct from other platforms by offering studies with multiple modalities, AI suggestions (e.g., in human-human studies, AI can suggest responses to a participant), live monitoring (e.g., researchers can evaluate, in real time, chats between communicators), and survey deployment (e.g., Likert-type scales, feeling thermometers, and open-ended text boxes can be sent to humans for in situ evaluations of the interaction), among other consequential features. No coding is required to operate Dyadic directly, and integrations with existing survey platforms are offered.

**Dyadic:**

**A Scalable Platform for Human-Human and Human-AI Conversation Research**

Conversation is the primary way in which people coordinate action, build relationships, and make meaning with others (Clark, 1996; Dunbar, 1996; Tomasello, 2008). The case for studying conversation often rests on at least three arguments: (1) conversation is a central mechanism through which most interpersonal outcomes of theoretical interest are produced, (2) conversation offers behavioral richness at timescales that other human behaviors may not, and (3) conversation has the unique capacity to reveal dynamic, emergent phenomena that individual-level approaches or behaviors cannot capture. Modeling and capturing conversation as an *interactive process* is inherently difficult, however, because of within- and between-person dependencies, measurement fidelity concerns, and other nontrivial issues (Solomon et al., 2023). Resolving such issues is critical for the development of our theories regarding how communication and language production or processing work. Indeed, dozens of perspectives have advanced the field's thinking on the nature of conversation in everyday life (Cappella, 1985; Clark, 1996; Grice, 1975; Solomon et al., 2023; Tomasello, 2008; Walther, 1996; Yeomans et al., 2023), but few tools have been up to the task of studying human-human and human-AI conversations with the precision and scalability required for modern day social science (see Hewes (2015) for an early treatment). Dyadic, a new conversation tool reviewed and presented in the current tutorial, was designed to relieve these tensions (and many more) in empirical conversation research, and its features are described below in terms of how they address key theoretical and practical barriers for interaction research.

The primary purpose of this tutorial is to advance conversation research in the social sciences by presenting a tool for its study. In providing this tutorial, there are five overarching goals: (1) to introduce a scalable and flexible platform for studying conversation, enabling researchers to capture the dynamic coordination between communicators; (2) to bridge traditionally siloed approaches that study interaction (e.g., those within computer-mediated communication, AI-Mediated Communication and psychology of language research) (Boyd & Markowitz, 2025; Hancock et al., 2020; Walther, 1996); (3) to provide researchers with the ability to manipulate conversational features (e.g., partner type, interlocutor identities) and embed in situ measurements (e.g., surveys) that test hypotheses about interaction processes; (4) to expand the scope of conversation research by enabling the study of human-AI and human-human interaction, which can offer new insight into how generative technologies modify social and psychological processes (Boyd & Markowitz, 2026), and (5) to allow researchers to generate granular-level analyses that can be linked to social and psychological outcomes, facilitating the study of how verbal behavior scales to broader interpersonal processes. This paper succeeds if it provides a brief and forward-thinking overview of Dyadic while sparking an interest in new scholars who may have been uninterested in conversation because the data were difficult to collect. It will also succeed if it interests scholars who may have given up on conversation research because tools and outputs fell behind user demand and expectations.

## A Platform Tutorial and Overview of Dyadic

Dyadic — https://www.chatdyadic.com/ — is a web-based research platform designed for the social scientific study of human-human and human-AI conversation. Dyadic is positioned as a stand-alone research platform and an integrative tool for use with existing participant recruitment sites. Simply put, researchers create accounts, configure studies and chatroom sessions, and export their data through a user dashboard. Dyadic is customizable, but it can be used "out of the box" by both amateurs and authorities in this space.

Dyadic is deployed on cloud infrastructure and is designed to allow concurrent sessions across many chatrooms. Real-time messaging is handled through persistent WebSocket connections, ensuring low-latency interaction regardless of geographic distance between participants. A visual overview of Dyadic is offered in Figure 1 alongside a description of its key functions in Table 1. Below, the sections that follow describe the platform's researcher- and participant-facing features that are relevant to study design and data collection.

**Organizational Structure of Dyadic**

Dyadic organizes research activities into a four-level hierarchy: researchers have accounts, accounts contain studies, studies contain rooms, and rooms contain participant slots. A study is an overall research project, while rooms are one fundamental unit of analysis. Each room hosts exactly one conversation session between two or more participants (currently, Dyadic allows up to 10 participants per room, but this can be adjusted based on researcher needs). Participant slots are fixed identities within a room (e.g., they are numbered sequentially and persistent across the entire session, but participant names can be modified for different purposes specific to a scholar's research question).

The participant slotting system is an important design feature of Dyadic. Given that each slot represents a specific person or role within an interaction, researchers can implement within-room manipulations that treat the two participants differently. A researcher studying persuasion, for instance, might configure Slot A as the message sender and Slot B as the recipient, providing each with different pre-session instructions about their role and goal. In a different study, a researcher studying social support might randomly assign Slot A to an advice-giver condition and Slot B to an advice-receiver condition. This framework enables designs that would be complex or impossible to achieve in most in-person laboratory paradigms using other tools.

*Study Types in Dyadic*

Currently, Dyadic offers two study types: (1) experimental and (2) observational. Between-subjects experiments can be implemented at the study level by creating room "sets," each configured with the relevant condition-specific parameters (e.g., all rooms in one set containing one AI communicator, all rooms in another set containing human-only communicators). Within a specific room, participants can also be randomly assigned to receive different prompts from the researcher before beginning the chat. Note, within-subjects experiments are also possible if one participant completes sessions in multiple rooms.

Observational studies do not assign rooms to conditions and instead, they operate as standard chatrooms. For example, a study seeking to evaluate how human-human or human-AI immediacy is achieved (e.g., Walther & Burgoon, 1992) can be initiated and completed with ease. Rooms can be created in bulk (e.g., dozens of rooms to be designated for a treatment condition, dozens of rooms to be designated for a control condition if an experiment is desired) for implementation on different research platforms. Importantly, data at the study and room levels can be exported, as rooms and participants are provided with unique identifiers.

**Conversation Modalities**

Dyadic supports text-based and real-time audio conversations between humans and humans and AI (see Figure 2 for screenshots of the text-based and audio-based interfaces). Participants can exchange text-based messages using an interface that displays the full conversation history, any relevant session information (e.g., timestamps), and, when applicable, embedded survey questions (see below). The session interface is deliberately minimalist, presenting only what the researcher has configured and avoiding visual clutter. Researchers can set a fixed session duration in seconds, and a countdown timer can be displayed to participants if desired. Researchers also can ensure that both participants have actively acknowledged their readiness before the conversation clock begins. Rather than starting the timer immediately upon both participants connecting, Dyadic holds the session in a "waiting" state until each participant clicks a confirmation button.

Pre-chat instructions are another critical control feature of Dyadic. Before a conversation begins, each participant can be shown a unique text block covering task instructions, a cover story, scenario descriptions, and so on. These instructions are slot-specific, meaning the two participants in a room can receive different information without any indication that their counterpart has seen something different. A researcher studying deception detection, for example, might trigger suspicion to the person in Slot A, while the person in Slot B is simply told to have a conversation.

Audio rooms allow voice-based interaction without requiring participants to install any software. Audio is transmitted through the browser's microphone API, making voice participation possible on any device that supports microphone access. From a participant's perspective, they would simply grant microphone access when prompted by the browser. The audio modality supports all the same session controls available for text-based rooms.

**Who Communicates with Whom? Human-Human and Human-AI Conversations**

One of Dyadic's central features is the ability to have an automated or AI-powered communicator in a conversation with other humans. AI participants occupy a specific slot within a room and are indistinguishable from human participants (unless the researcher decides to make this clear by naming the participant Chatbot, etc.). A researcher can designate when AI-generated messages are pushed using a configurable response delay. That delay can be set as a fixed interval (e.g., 2,000 milliseconds after the human response) or as a random draw between time intervals (e.g., 2,000 and 4,000 milliseconds). This feature allows researchers to control or manipulate response latency, which is often a research interest for scholars studying human-human conversation (Kalman et al., 2006; McLaughlin & Cody, 1982), and is a pressing concern to make human-AI conversational feel appropriate (Kim et al., 2025). Response delay can also be an experimental manipulation, as a researcher studying rapport building might examine how a person who responds slowly affects the conversation quality versus a person who replies quickly.

Text-based AI bots are powered by large language model APIs (API keys are stored, securely, in each researcher's settings). Currently, access to four APIs is offered via OpenAI, Anthropic, Google Gemini, and Hugging Face. In conversations with an LLM, researchers decide on the models to use and provide a system prompt that defines the AI's persona, communicative role, topic knowledge, and so on. For voice-enabled rooms, researchers can configure the AI's voice, a threshold that determines when the system interprets a pause, and a microphone input mode of either always-on (e.g., the AI listens continuously and responds whenever silence is detected) or push-to-talk (e.g., the participant must hold a button to speak).

**Researcher Interventions: AI Suggestions, Monitoring, and Human Surveys**

Dyadic provides three ways that researchers can intervene in a live conversation, if desired. The first is via AI Suggestions, where Dyadic allows a certain respondent to receive suggestions on how to reply to the other communicator(s) in the room. Dyadic currently instructs an LLM to account for (up to) the last 20 messages communicated in the chat to form three total suggestions. Users can click on one of the suggestions, which then appears in the dialogue box for participants to edit if desired before sending the message (see Figure 3). Participants in the same chat who are not assigned to have AI suggestions enabled will not see the suggestions.

The second intervention is live monitoring and direct message injection when a chat is live. Researchers can "live monitor" a chat — like a researcher would do in a "control room" in the lab —and they can inject messages (e.g., additional instructions, feedback to participants) if desired. This feature may be critical to ensure participants are on task during a chat, and to monitor how communicators are interacting when sensitive topics are discussed. Institutional Review Boards may be concerned when humans communicate with AI because it is unclear what topics might be raised by a LLM, and this feature allows a human to remain in the loop.

Finally, and perhaps the most innovative in situ offering by Dyadic, is its survey delivery feature. In most research designs, relational measures are taken after the conversation ends because questions regarding how perceptions of the interaction unfolded over time had been difficult or impossible to obtain. A more precise and granular way to evaluate how people think and feel about a conversation, their partner, or their experience is to measure these dynamics *throughout the conversation*. Dyadic allows researchers to embed surveys directly within a chat session. When researchers decide to trigger survey questions such as Likert-type scales, feeling thermometers, or open-ended questions, they appear in the same browser window (see Figure 4 for an example). With Dyadic, scholars no longer need to rely on participant recall at the end of an interaction to gauge how interactants thought and felt about the conversation.

Multiple surveys can be distributed within a single session and crucially, researchers can select when people are presented with such survey questions. Currently, the survey dissemination options for researchers are: (1) manually, or when a researcher submits a question to be answered; (2) automatically, after a specific number of seconds from the room start; (3) automatically, after a specific number of messages appeared in the room; (4) recurring, where questions are repeated after a certain number of seconds, and (5) after the chat ends. Participants have ten seconds to answer survey questions (the amount of time can be configured to a researcher's liking), and if responses are not submitted within that time allotment, the last known value is recorded. Every survey response can be linked to a person and to the specific conversation data that preceded it, a design that pairs naturally with dynamic systems analytic approaches (Solomon et al., 2023). Researchers can configure surveys through the study dashboard and build questionnaires from a question library (questions can be saved for future use). Surveys can be assigned to specific rooms or across all rooms in a study.

**Data Collection, Data Export, and Data Security**

Dyadic logs conversation data automatically and continuously throughout every session. Each message is stored with a millisecond-precision timestamp, the room identifier, the sender's slot position, the sender's display name, and a flag indicating whether the message was generated by a bot or a human participant. For text-based sessions, the full message-level transcript is preserved. For audio rooms, Dyadic provides a text transcript of the voice session to the database in the same format as text chat data. Researchers can display this transcript to participants in real time, but by default it is hidden.[1] All transcriptions are handled server-side via the OpenAI API. Dyadic uses *Whisper-1* for participant speech and the *gpt-4o* Realtime model's native transcript output for AI audio replies.

Export of all data, regardless of conversation modality, is available at the room and study levels in flat-file formats. Researchers are offered: (1) a Chat CSV, which contains the full chat log, and (2) a Survey CSV, which contains participant responses if surveys appeared during the session. Study-level exports aggregate all rooms into a single file per type. Session metadata (e.g., session start time, end time, duration, room condition labels, and participant slot assignments) are included in both export types, along with other data like response latency, typing behavior, and mouse clicks during message composition in case they are useful to researchers and their empirical questions (see Table 1 for full descriptions of these metrics).

---

[1] Researchers who need the audio for additional coding will need to supplement Dyadic with other recording tools.

Finally, Dyadic takes data security seriously. Researcher passwords are hashed using bcrypt prior to storage and API keys are encrypted at rest using AES-256-GCM. Participant IP addresses are never retained directly. All data transmission occurs over HTTPS with HTTP Strict Transport Security (HSTS) enforced at the server level, and authentication endpoints are protected by rate limiting and automatic account lockouts to mitigate unauthorized access attempts. Researchers access only the studies and rooms they own or have been explicitly invited to collaborate on, enforced at the database query level, meaning no researcher can inadvertently or deliberately retrieve another researcher's participants or data. Data exports are also authenticated and rate-limited to prevent systematic scraping. These measures are designed to meet baseline expectations for social science research and to provide researchers with reasonable assurance that participant confidentiality is maintained. The author of Dyadic has made a commitment to maintaining best practices for data and participant security.

## Limitations

Dyadic was designed for research settings where controlled, dyadic or small-group conversations, are the primary interest. The room structure, the slotting system, and the monitoring interface are all functional for dyads or small groups, and scaling to dozens of participants in a single session may require changes to the system. Further, audio rooms introduce dependencies that text sessions do not (e.g., browser microphone permissions must be granted by the participant). Users should therefore test how Dyadic works with research assistants prior to deploying with actual participants. Currently, Dyadic is also best suited for desktop studies; hopefully, future iterations of the tool will build out a phone-specific interface.

That is, the mobile experience has not been fully optimized for small-screen interaction. Researchers intending to recruit mobile participants should conduct pilot tests to ensure the participant experience is functional and parsimonious.

There is an important access limitation to Dyadic's AI capabilities. Obtaining an API token is free but using the API costs money for certain platforms. Dyadic can connect with Hugging Face and open-source models, which may relieve this constraint, but it is important to note that access is not inherently free for API use. On a complementary note, changes to the APIs or models could limit replicability or reproducibility (Gundersen & Kjensmo, 2018; Haibe-Kains et al., 2020). This is not a concern that is unique to Dyadic, but it is an important one to acknowledge, nonetheless. When using and publishing studies run on Dyadic, researchers should record model versions in their method sections and, where possible, to archive bot responses for study-by-study comparison.

## Conclusion

Dyadic was built to lower the barriers to entry for conversation research between humans and between humans and AI. By providing a scalable and flexible platform, conversation studies are now more accessible and attainable for researchers who have longed to study such interactions but have lacked the practical means to execute them. The timing of this platform's creation is also not happenstance. The emergence of LLMs as conversational partners has created needs and opportunities for a tool to study social and psychological dynamics of these different interlocutors. Dyadic is specifically designed to study human and non-human partner types across different communication modes (text vs. audio), which introduces scholars with generative possibilities for their emerging research questions.

**Author's Note**

Dyadic was conceptualized and written by the author. Initial underlying code was written by him and significantly augmented by Claude Code. Dyadic is a living tool and researcher suggestions are welcomed for improvements to the platform. Please contact the author if you notice bugs, improvements, or have specific functional needs that you would like implemented.

**Table 1**

*Functional Overview of Dyadic*

| Category | Feature | Description | Modality | Study Type | Researcher-Facing | Participant-Facing |
|---|---|---|---|---|---|---|
| **Study Structure and Organization** | | | | | | |
| | Researcher accounts | Researchers create accounts to manage their own data. | Both | Both | Yes | No |
| | Study hierarchy | Four-level hierarchy: (1) researcher → (2) study → (3) room → (4) participant slot. | Both | Both | Yes | No |
| | Study types | Experimental (condition assignment, manipulations) or Observational (chatrooms with no condition structure). | Both | Both | Yes | No |
| | Participant slots | Each room has numbered, fixed slots (up to 10). Slots persist across the session and enable slot-specific configurations. Participant names are customizable. | Both | Both | Yes | Yes |
| | Bulk room creation (CSV import) | Researchers import a CSV to create dozens of rooms at once, with optional condition labels and configurations. | Both | Both | Yes | No |
| | Collaborator/team access | Studies can be shared with collaborators who gain access to rooms and data exports. | Both | Both | Yes | No |
| | External platform integration | Dyadic generates unique participant URLs compatible with participant recruitment platforms. | Both | Both | Yes | No |
| **Conversation Modalities** | | | | | | |
| | Text-based chat | Real-time typed messaging between participants. Full message history displayed in browser. | Text only | Both | No | Yes |
| | Audio/voice-based chat | Real-time voice interaction via browser microphone API. Works on any microphone-enabled device. | Audio only | Both | No | Yes |
| | Session duration timer | Researcher sets a fixed session length in seconds. Countdown can optionally be displayed to participants. | Both | Both | Yes | Yes |
| | Ready confirmation gate | Session clock does not start until all participants have clicked a confirmation button, preventing premature start. | Both | Both | Yes | Yes |
| | Pre-chat instructions (slot-specific) | Each slot can be shown a unique text block (e.g., task instructions, cover story) before the conversation begins. | Both | Both | Yes | Yes |
| **Partner Type & AI Participation** | | | | | | |
| | Human–human conversations | Two or more human participants connect to the same room and interact in real time. | Both | Both | No | Yes |
| | Human–AI text bot | AI agent occupies a participant slot, powered by an LLM API. Researcher provides system prompt. | Text only | Both | Yes | Yes |
| | Human–AI voice bot | Voice-enabled AI agent in an audio room. Researcher configures prompt, silence threshold, and microphone input mode. | Audio only | Both | Yes | Yes |
| | Bot response delay (fixed) | AI text bot waits a fixed number of milliseconds before replying. | Text only | Experimental | Yes | No |
| | Bot response delay (random range) | AI text bot draws a random delay from a researcher-specified min–max range. | Text only | Experimental | Yes | No |
| | Bot system prompt / persona | Researcher writes a system prompt defining the AI's persona, role, etc. | Both | Both | Yes | No |
| | Voice input mode (always-on / push-to-talk) | Audio rooms support continuous listening or push-to-talk modes for voice AI interaction. | Audio only | Both | Yes | Yes |
| **Researcher Interventions** | | | | | | |
| | Live session monitoring | Researchers view a live feed of any active room from the monitor dashboard. | Both | Both | Yes | No |
| | Message injection | Researcher manually injects a message into a live chat (e.g., to provide additional instructions, feedback). | Both | Both | Yes | No |
| | AI suggestions | Up to 3 AI-generated reply suggestions based on the last 20 messages. Clicking a suggestion pre-fills the input box. | Text only | Experimental | Yes | Yes |
| | AI suggestions trigger (every message) | Suggestions are generated and shown after every message received by the target participant. | Text only | Experimental | Yes | No |
| | AI suggestions trigger (every N messages) | Suggestions are generated every N messages (configurable count), reducing frequency of AI prompts. | Text only | Experimental | Yes | No |
| | AI suggestions trigger (manual) | Target participant can request suggestions on demand by clicking a button. | Text only | Experimental | Yes | Yes |
| **Survey Delivery** | | | | | | |
| | Embedded survey delivery | Survey questions (Likert scales, feeling thermometers, and open-ended text boxes) appear in the same browser window as the chat. | Both | Both | Yes | Yes |
| | Survey trigger: manual | Researcher pushes a survey question to participants in real time from the monitor interface. | Both | Both | Yes | No |
| | Survey trigger: after N seconds | Survey appears automatically after a specified number of seconds from room start. | Both | Both | Yes | No |
| | Survey trigger: after N messages | Survey appears automatically after a specified number of messages have been sent in the room. | Both | Both | Yes | No |
| | Survey trigger: recurring | Survey repeats at a fixed time interval (e.g., every 60 seconds) throughout the session. | Both | Both | Yes | No |

| | | | | | |
|---|---|---|---|---|---|
| Survey trigger: post-chat | Survey appears after the conversation ends. | Both | Both | Yes | No |
| Survey auto-submit timer | If participants do not submit a survey within a set time, the last known value is submitted automatically. | Both | Both | Yes | No |
| Question library | Researchers save questions for reuse across studies. | Both | Both | Yes | No |
| Bulk survey assignment | Surveys are assigned to a single room or propagated across all rooms in a study at once. | Both | Both | Yes | No |
| **Experimental Design and Manipulation** | | | | | |
| Between-subjects conditions | Distinct room sets are each configured with condition-specific parameters for between-subjects experiments. | Both | Experimental | Yes | No |
| Within-room slot manipulations | Each slot receives different instructions, prompts, or conditions within the same room. | Both | Experimental | Yes | No |
| Prompt/condition randomization (study pool) | Researcher defines named conditions with slot-specific text; system randomly assigns a condition to each room. | Both | Experimental | Yes | No |
| Prompt/condition randomization (shuffle) | Existing slot texts are shuffled within a room using a single random permutation to keep labels paired. | Both | Experimental | Yes | No |
| **Data Collection** | | | | | |
| Chat log (text) | Every message stored with millisecond-precision timestamp, room ID, slot, display name, and human/bot flag. | Text only | Both | Yes | No |
| Chat log (audio transcript) | Voice sessions are automatically transcribed server-side (OpenAI Whisper-1 for participants; gpt-4o Realtime for AI). Stored in same format as text logs. | Audio only | Both | Yes | No |
| Response latency — first keystroke | Time from message delivery to recipient's first keystroke logged per message. | Text only | Both | Yes | No |
| Response latency — reply send | Time from message delivery to when the recipient sends their reply. | Text only | Both | Yes | No |
| Typing behavior | Total typing duration, keystroke count, edit/deletion count, and paste count per message. | Text only | Both | Yes | No |
| Mouse clicks | Click count recorded during message composition. | Text only | Both | Yes | No |
| Survey response linkage | Every survey response is linked to the participant, room, session metadata, and the conversation data that preceded it. | Both | Both | Yes | No |
| **Data Export** | | | | | |
| Room-level Chat CSV export | Flat-file export of the full chat log for a single room. | Both | Both | Yes | No |
| Room-level Survey CSV export | Flat-file export of all survey responses for a single room. | Both | Both | Yes | No |
| Study-level Chat CSV export | Aggregated chat log across all rooms in a study in a single file. | Both | Both | Yes | No |
| Study-level Survey CSV export | Aggregated survey responses across all rooms in a study in a single file. | Both | Both | Yes | No |
| Session metadata in exports | Start/end time, duration, condition labels, and slot assignments included in all CSV exports. | Both | Both | Yes | No |
| **Data Security** | | | | | |
| Privacy and security | Researcher passwords are hashed using bcrypt prior to storage; API keys are encrypted at rest using AES-256-GCM. | Both | Both | Yes | No |
| IP address hashing | Participant IP addresses are never stored directly; hashed for abuse prevention only. | Both | Both | Yes | No |
| HTTPS / HSTS | All data transmission encrypted via HTTPS; HTTP Strict Transport Security enforced at server level. | Both | Both | Yes | No |
| Rate limiting & account lockout | Authentication endpoints are rate-limited; accounts lock out after repeated failed attempts. | Both | Both | Yes | No |
| Researcher data isolation | Each researcher accesses only their own studies/rooms, enforced at the database query level. | Both | Both | Yes | No |
| Export rate limiting | Data exports are authenticated and rate-limited to prevent systematic data scraping. | Both | Both | Yes | No |

19**Figure 1**

*Overview of Main Dyadic Features*

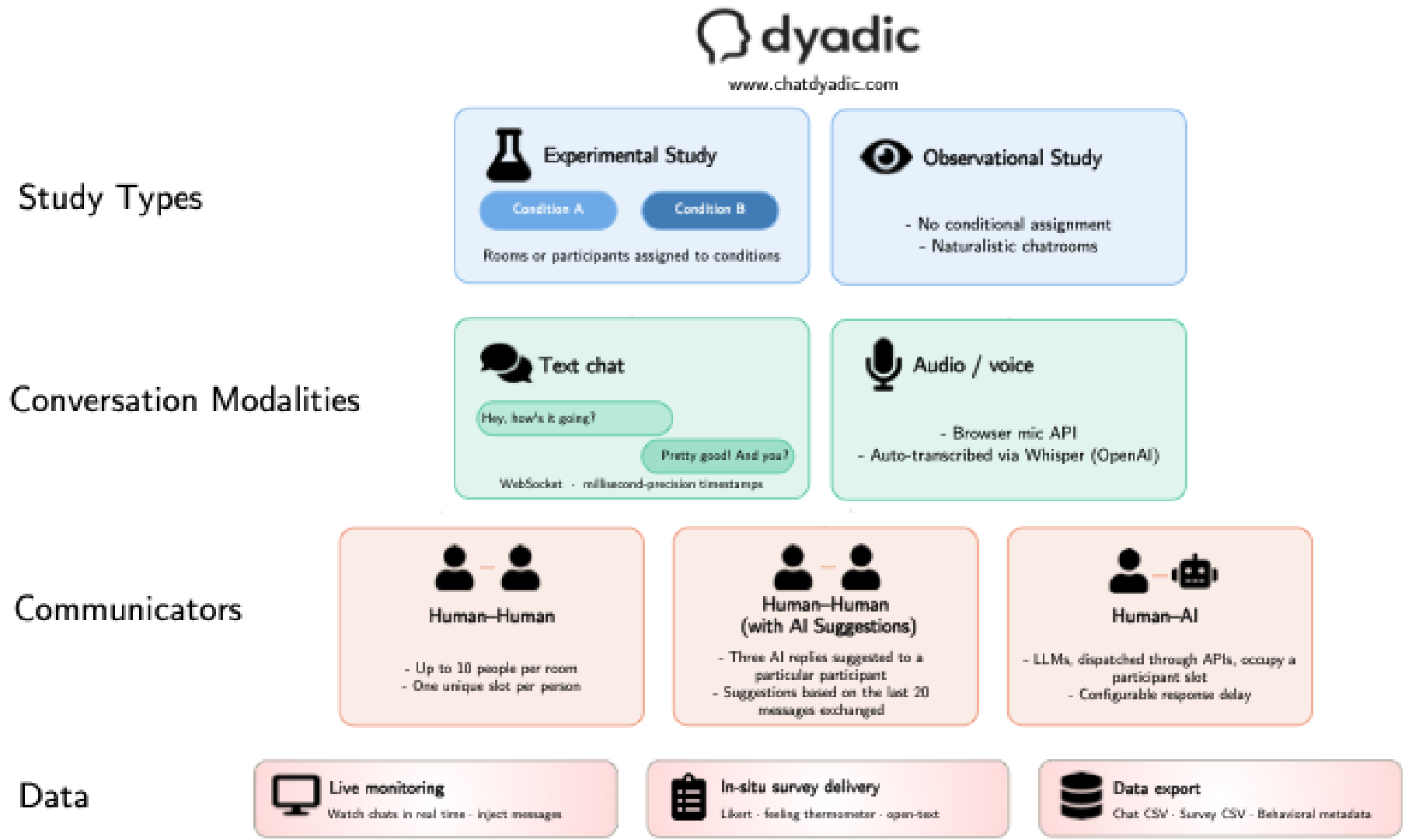

**Figure 2**

*Modality Differences Between Text and Audio*

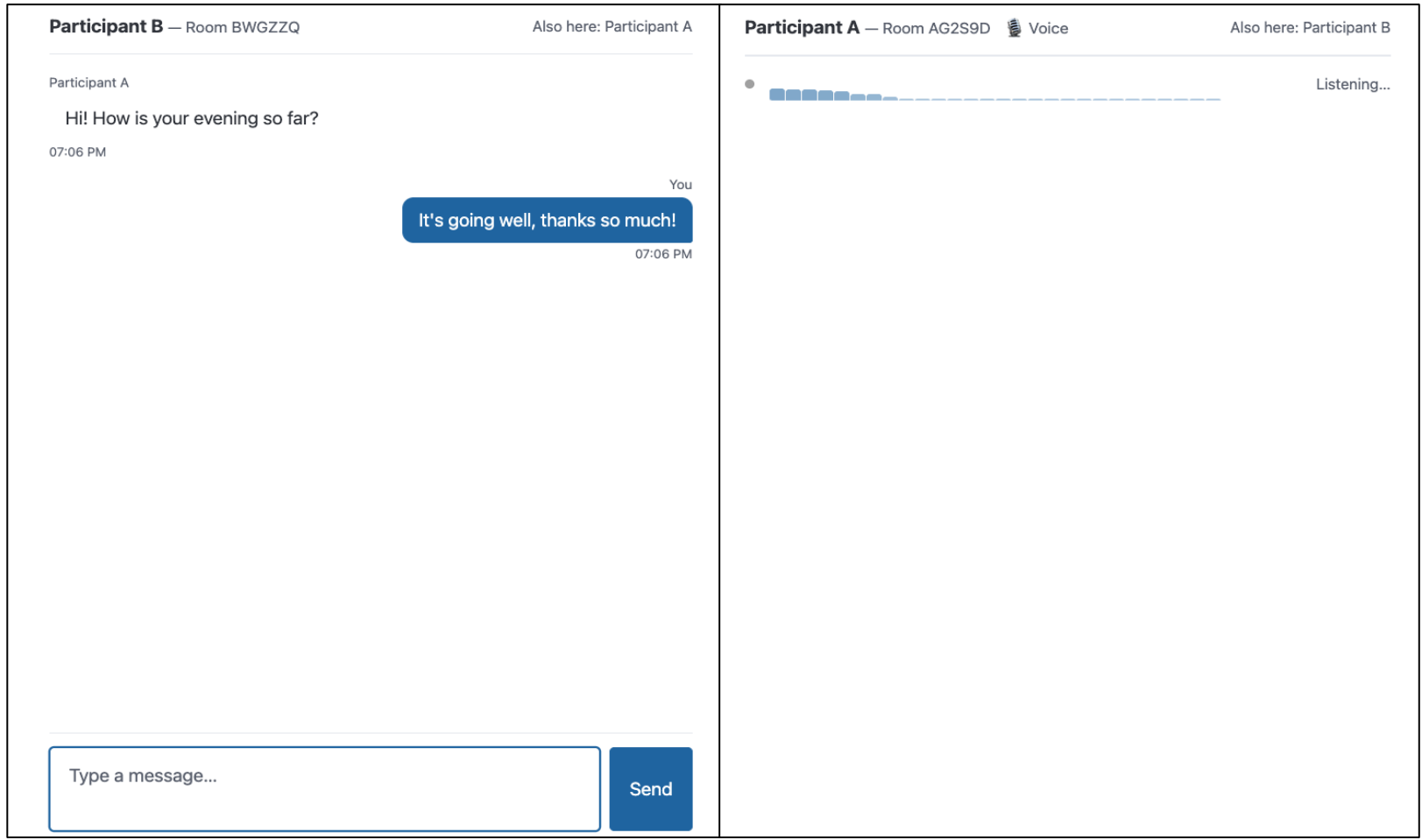

*Note*. The left panel represents a text-based chat, and the right panel represents an audio chat in Dyadic. Bars in the audio chat indicate sound being received from the audio system while a person is communicating.

**Figure 3**

*The AI Suggestion Feature in Dyadic*

*Note*. AI suggestions must be enabled to appear. At most three responses are displayed (left panel). When one is clicked (right panel), it appears in the dialogue box, where participants can edit the message before submitting to the other chat partner.

22**Figure 4**

*Visual Description of the Survey Feature on Dyadic*

*Note.* A feeling thermometer (left panel) and Likert-type scale question (right panel) on Dyadic embedded in the chat.